\documentclass[structabstract]{aa}  
\usepackage{graphicx}
\usepackage{txfonts}
\usepackage{natbib}
\begin{document}
   \title{In the Neighbourhood of Tame Monsters.}

   \subtitle{A study of galaxies near low-redshift quasars.}

   \author{Beatriz Villarroel
          \inst{1}
	  }

   \institute{Department of Physics and Astronomy\\
		Uppsala Universitet\\
              SE-751 20 Uppsala, Sweden\\
              \email{beatriz.villarroel@physics.uu.se}
        }

   \date{Received November 22, 2011; accepted into A\&A April 23 2012}

% \abstract{}{}{}{}{} 
% 5 {} token are mandatory
 
  \abstract
   {The impact of quasars on their galaxy neighbours is an important factor
   in the understanding of galaxy evolution models.}
   {The aim of this work is to characterize the intermediate-scale environments of quasars at low redshift (z $<$ 0.2) with the most
   statistically complete sample to date using the seventh data release of the 
   Sloan Digital Sky Survey.}
   {We have used 305 quasar-galaxy associations with spectroscopically
   measured redshifts within the projected distance
   range of 350 kpc, to calculate how surface densities of galaxies,
   colors, degree of ionization, dust extinction and star-formation rates change as a function of the distance
   to our quasar sample. We also identify the companion Active Galactic Nuclei from our main galaxy sample 
   and calculate surface density for different galaxy types.
   We have done this in three different quasar-galaxy redshift difference ranges   
    $|\Delta$z$|<$ 0.001, 0.006, and 0.012.}
   {Our results suggest that there is a significant
   increase of the surface density of blue neighbours around our low-redshift quasar sample that is steeper than around non-active field galaxies of the same luminosity and redshift range. This
   may indicate that quasar formation is accomplished via a merging scenario. No significant changes in star formation rate, dust extinction, degree of ionization or color as a function of distance from the quasars was observed. We could not observe any direct effects from quasars on the their companion galaxies.}
   {}

   \keywords{active --
                nuclei --
                star formation --
		quasars: general --
		interactions --
		evolution
               }

   \maketitle
%
%________________________________________________________________

\section{Introduction}\label{sec:intro}

Since their discovery in 1963 \citep{Schmidt1963} quasars have sparked widescale
interest among astronomers. Their large luminosities and variation over redshift 
are incomparable to that of any other galaxy type, while their decline in number in our 
Local Universe remains poorly understood. Low-redshift quasars are not only under-represented, 
but also considerably fainter compared to their high-redshift members. Regardless of whether they were 
victims of their own high-energy consuming nature, or rather strangled their less luminous surrounding galaxies, traces of their dynamic history could be found by observing in which way these most extreme active galactic nuclei influence the evolution of their galaxy neighbours. These effects could mark either direct influence of AGN, or trace the environments most favorable to their occurrence.

The extremely luminous nature of quasars is believed to be driven by an active galactic 
nucleus due to accretion of material upon a supermassive black hole \citep[e.g.][]{LyndenBell1969,Rees1984}. 
Feeding such a black hole may affect the formation of galaxies in proximity to quasars 
\citep{Gunn1979,SR1998}, and this could play a large role in structure formation scenarios derived from 
standard cosmological models.

Could quasars be the result of mergers according to a hierarchical structure formation scenario 
predicted by Cold Dark Matter Models? For many years there have been indications that 
both nuclear activity and star formation may be triggered or enhanced by interactions between galaxies. 
For example, the presence of post-starburst populations in close quasar companions \citep[e.g.][]{Canalizo1997, Scannapieco2005, Almeida2011}, morphological asymmetries with high star formation rates near quasar hosts \citep{Gehren1984,Hutchings1984} and possible quasar companion galaxies \citep{Canalizo2001}.
Ionized gas has also been observed around quasars \citep[e.g.][]{Boroson1985}, while others studies have suggested that high-luminosity quasars could ionize the galactic medium up to several megaparsecs away \citep{Rees1988,Babul1991}.
	
A number of additional studies have suggested that quasars influence the evolution of normal galaxies. 
There are essentially two different populations of normal ``non-active" galaxies; 
red mostly quiescent galaxies with early-type morphology, and blue star-forming galaxies
\citep{Strateva2001,Brinchmann2004}. The color bimodality of these galaxy populations suggest that they originate from a mechanism that quenches star formation in the blue galaxy population turning them red.
This hypothesis has resulted in a number of ideas of possible feedback mechanisms associated with active
galactic nuclei \citep[e.g.][]{TB1993,Hopkins2006,Sijacki2006}. These feedback mechanisms
have been proposed as an alternative to more conservative explanations such as
mergers between late-type galaxies 
\citep[e.g.][]{Toomre1972,Barnes1992} or loss of gas due to ram pressure stripping \citep{Gunn1972}.

However, more study is needed in order to distinguish the various scenarios presented above.
A great deal of work has gone into the study of close pairs (less than 30 kpc) of galaxies \citep[e.g.][]{Rogers2009} and of the surrounding environments of quasars at larger radii (larger than 1 Mpc) most recently by Coldwell \& Lambas (2006) hereafter CW06, but there have been few detailed studies on an intermediate scale. In the previous larger-scale study of quasar environments
up to 3 Mpc by CW06, a small (20\%) excess of blue star-forming galaxies around quasars was shown for redshifts z $<$ 0.2. CW06 also demonstrated that quasars tend to avoid high-density environments such as clusters and galaxy groups. This conclusion has been supported by other studies \citep[e.g.][]{Lietzen2009}. Active Galactic Nuclei (AGN) also tend to avoid high-density environments 
\citep[e.g.][]{Way1998,Arnold2009}. 

Many studies of quasar environments using the Sloan Digital Sky Survey (SDSS)
have used galaxies with photometrically determined redshifts (photometric redshifts, Photo-Z)
which have large redshift uncertainties $\delta_{z} >$ 0.025. These allowed more quantitative studies than qualitative, which underlines the importance of careful quasar-galaxy association studies with spectroscopically defined redshifts with small errors $\delta_{z} <$ 0.001.

In the CW06 investigation SDSS quasars were restricted to redshifts $z <$ 0.2 and luminosities brighter than M$_{r}$ $<$ -23. The aim was to explore the distributions of quasars in different environments. They also examined the distribution of the
quasar colors and magnitudes, as well as the morphology and spectral type of the galaxies surrounding the 
quasars. However, their results were 
difficult to interpret due to the fact that they used fraction of total
number of galaxies in each histogram bin without informing the reader of the total number
of galaxies per bin. They also relied on the SDSS spectral classification coefficient (eClass)
to estimate the morphology of the quasar companion galaxies. The eClass coefficient is calculated
from Principal Component Analysis (PCA). We show in Section \ref{sec:morphology} that this method is not an accurate classification of morphology.

We use a non-volume-limited sample drawn from the seventh data release (DR7) of the SDSS.
The DR7 contains more than 100,000 quasars of which we use a small subsample to study how quasars influence their nearby galaxies at low redshifts. 
We focus on examining how colors, surface densities, star formation rates and 
ionization of the galactic medium of galaxies near quasars change depending on the distance between the quasar and the neighbour.
This is done for a sample consisting of 305 quasar-galaxy associations, at redshifts $z <$ 0.2 using Standard Cosmology
($\Omega_{\Lambda}$=0.70, $\Omega_{M}$=0.30, H$_{o}$=70 km s$^{-1}$ Mpc$^{-1}$). In a future publication
we plan to replicate this study for higher-redshift objects to investigate possible redshift evolution 
in quasar-galaxies associations. With these two studies,
it may be possible to understand how quasars impact surrounding galaxies as a function of time.
Such a comprehensive study could better constrain AGN feedback models and contribute to the understanding of the formation of the most luminous galaxies in our Universe.

In Section \ref{sec:2} we describe our sample selection, Section \ref{sec:3} will discuss how we
estimate the environment around the quasars, Section \ref{sec:4} is a discussion
of the results of our analysis and conclusions.

%__________________________________________________________________

\section{Sample Selection}\label{sec:2}

\subsection{Spectroscopic redshift sample}\label{sec:2.1}

The quasar and galaxy samples were drawn from the SDSS \citep{York2000} DR7 \citep{Abazajian2009}. 
The spectroscopic data in this catalogue contains 929,555 galaxies and 121,363 quasars in 
five optical bands (u, g, r, i, z) that cover the complete wavelength range of the CCD-camera.
	
The quasars were extracted from the SDSS quasar catalogue \citep{Schneider2003}. These have been
selected such that they have at least one emission line with full width at half maximum 
(FWHM) larger than 1000 km s$^{-1}$ and $M_{i}$ $<$ -22, and have highly reliable redshifts ($zConf >$ 0.95). Both the 
quasars and galaxies were chosen to be in the redshift range 0.03 $<$ z $<$ 0.2 which means that most of
the quasars we have are low luminosity quasars and Seyferts rather than conventional high-luminosity quasars.
The limited apparent luminosity of the spectroscopic selection of quasars implied by Schneider et al. (2002) will likely help us to avoid the low luminosity Seyfert 1s. Objects were excluded from our sample if they were flagged via the SDSS photometric pipeline \citep{Lupton01} as overly bright (flags\&0x2=0), saturated (flags\&0x40000=0), or from blended images (flags\&0x8=0) \citep[see Table 9]{Stoughton02}. We do a SDSS query for quasars within the redshift range to see how many we get with and without the saturation, blending and brightness flags. We observed a loss of 5\% of the quasars, an effect that we consider being negligible. As quasars are the most luminous among AGN, we do not expect this to influence the number of AGN neighbours later in our analysis. We used the BestObjIDs in our queries for both galaxies and quasars, since the imaging data in BESTDR7 is of the highest quality\footnote{See the SDSS casjobs Schema browser for more information http://casjobs.sdss.org/dr7/en/help/browser/browser.asp}. This is particularly important for the careful selection of blue galaxies in this study.

To obtain the quasar neighbours the casjobs\footnote{http://casjobs.sdss.org} query was done in two parts: a low redshift part 
0.03 $<$ z $<$ 0.1 where we searched for neighbouring galaxies up to 10$\arcmin$ and a high-redshift part 0.1 $<$ z $<$ 0.2 
with companions within 7$\arcmin$ of the quasar. We limited the redshift difference between the galaxy and
the companion to be $|\Delta$z $|<0.012 $ in the queries so that this may be compared to similar Photo-Z studies which
have photometric redshift errors of this order. The goal is to obtain any galaxy within 350 kpc of a quasar
in our sample.
%The queries were constructed to avoid having a sample with a biased number distribution of 
%galaxies within the projected distance range that is most in our interest in the later analysis, which is between 0 and 350 kpc.
With the minimum 55$\arcsec$ fiber separation, pairs closer than 100 kpc from each other at redshifts z $\ga$ 0.1 will not
be measured on the same spectroscopic plug plate. Many of the pairs with r $<$ 17.7 will be seen in the SDSS imaging
survey, but will lack a redshift measurement for one of the objects. For both quasar-galaxy pairs within 100 kpc and with redshifts z $\ga$ 0.1
to be spectroscopically measured they have to reside in plate overlap regions. For this reason we have a particularly small sample size.

The following quantities were retrieved for each quasar and galaxy in our sample:
\begin{itemize}
\item spectral line (absorption or emission) information
\item apparent dereddened magnitudes (corrected for extinction)
\item K-corrections and rest-frame absolute magnitudes
\item isophotal axis length
\item radii including 50\% and 90\% of the Petrosian flux in the i-band
\end{itemize}

We calculated the projected distances for all 305 quasar-galaxy associations (there were 284 unique quasars).
Galaxies with internal extinction corrected colors of $U_{e}-R_{e}<2.2$ (using rest-frame magnitudes)
will henceforth be referred to as ``blue" galaxies \citep{Strateva2001}, while the remaining
galaxies will be called red (see Section \ref{sec:3.2}).

In subsequent calculations of surface densities, star-formation rates, ionization degree and colors we have separated AGN from normal galaxies (see Section \ref{sec:2.3}).

\subsection{Line flux and extinction corrections}\label{sec:2.2}

The Balmer emission line fluxes and equivalent widths (EW) were corrected for underlying stellar 
absorption by assuming average absorption line strengths corresponding to 2.5 \AA\ 
in $EW$ for H$\alpha$ and 4 \AA\ for H$\beta$. Internal extinction corrections were only done if 
$EW$(H$\beta$) $>$ 5 \AA\, since H$\beta$ lines lower than 5 \AA\ are too noisy
to permit a reliable extinction correction. The spectral lines were internal extinction-corrected 
following a standard interstellar extinction curve \citep[e.g.][]{Whitford,Osterbrock2009}.

The colors were corrected for extinction by taking the inclination of the observed galaxies into account.
The was done for all galaxies fulfilling the criteria described in \cite{CP2009} using
their derived analytical expressions. Colors were corrected for galaxies having
0 $<$ u-r $<$ 4, H$\alpha$ 0-200 \AA\, absolute magnitude -21.95 $ < M_{r} < $-19.95 
and a concentration index (CI) in the range 1.74 $< CI <$ 3.06.

At earlier stages of our analysis, we did the same study without any line flux corrections. The addition of line flux corrections has not influenced the conclusions.

\subsection{AGN activity}\label{sec:2.3}

In some theories AGN can be responsible for inducing star-formation activity in nearby galaxies, while in other theories they are hypothethized to quench star formation via AGN feedback. In order to investigate either of these possibilities it is necessary to separate AGN from normal star-forming galaxies in the quasar companion sample. 
In order to separate AGN from starforming \ion{H}{ii} type galaxies in our neighbour galaxy sample we used BPT line-ratios \citep{BPT1981} combined with \cite{Kauffmann2003} criteria:

\begin{equation}
log([\ion{O}{iii}]/H\beta) > 0.61/(log([\ion{N}{ii}]/H\alpha))-0.05)+1.3
\end{equation}\label{eq:BPT}

To remove the remaining broad-line (type 1) AGN an additional criterion was added, $\sigma ($H$\alpha)>15$ \AA. With these two criteria 69 AGN were identified in the companion galaxy sample.

\subsection{Field galaxies}\label{sec:2.4}

A companion field galaxy sample was obtained from the papers of Way et al. (2011, in prep) 16526 galaxies were classified as ``field" (Iblock=12--19) by the ``Bayesian Blocks" method outlined in the aforementioned papers. 16509 of the original ``field" sample had SDSS casjobs Object IDs. The object IDs were necessary to obtain additional photometric and spectroscopic parameters not present in the original catalogs. Among these we selected non-AGN field galaxies using the criteria discussed in Section \ref{sec:2.3}. Using the same procedure outlined in Section \ref{sec:2.1} to obtain neighbours for the quasar sample, we obtained 435 field galaxy-galaxy pairs in the redshift range 0.03 $<$ z $<$ 0.1.

\subsection{Photometric redshift samples}\label{sec:2.5}

Approximately 1/3 of the Sky is covered by more than one spectroscopic plate, and it is estimated that roughly 2/3 of all pairs with angular separation less than 55$\arcsec$ will not be detected. This may bias
our sample towards more wide pairs. To examine how the conclusions on eventual clustering around quasars can be influenced by the fiber collision constraint, we constructed a photometric redshift sample with neighbours from the Photo-z catalogue. For the quasars already existing in our spectroscopic quasar-galaxy pair sample, we searched for all neighbouring galaxies within the projected distance of 350 kpc and $|\Delta$z$|<0.03 $ from the Photo-z catalogue. We also did the same for our central field galaxies in the spectroscopic field pair sample. As a result, we obtained in total 3127 quasar-galaxy pairs and 4214 field galaxy-galaxy pairs.

\section{Results}\label{sec:3}

\subsection{Morphology classification}\label{sec:morphology}\label{sec:3.1}

The companion galaxies in the sample underwent a morphological classification in order to make a 
comparative analysis between disk type and spheroidal galaxies.

The first method attempted was based on a Principal Component Analysis (PCA) 
of spectroscopic objects using the method of \cite{CS1999}.
The spectral classification coefficient is generally calculated by using 
two out of five eigencoefficients that build up the expansion of the eigentemplates of a 
galaxy's spectrum. These eigencoefficients can be extracted from the SDSS catalog and are defined as:

\begin{equation}
eClass= \mathrm{arctan} (-eCoeff2/eCoeff1).  
\end{equation}

Early-type galaxies have an eClass value -0.35 $ <$ eClass $< 0$, while late-types have 0 $<$ eClass $<$ 0.55 \citep{CL2006}.

Using the SDSS casjobs interface we obtained the eCoefficients eCoeff1 and eCoeff2 for
the 305 quasar companion galaxies.  Out of the 305 galaxies 183 were defined by their eClass as late-type, 34
as early-type, while the remaining 88 did not fit the eClass criteria.

\onecolumn
\begin{table*}[ht]
\caption{Visual inspection of 183 neighbour galaxies detected by eClass as ``late-type". The right column shows the results from the visual inspection of these eClass-defined
late-type galaxies. The 41 remaining early-type galaxies predicted from eClass, did not undergo
any visual inspection and are therefore not included in the table.}
\centering
\begin{tabular}{c c c}
\multicolumn{3}{c}{eClass morphology classification} \\
\hline\hline
Galaxy type & eClass results & Visual inspection \\ %[7ex]
	& (\#) & (\#)\\
\hline
Total & 183 & 183 \\
Late-type & 183 & 70 \\
Early-type or indefinable & 0 & 113 \\ [1ex]
\hline
\end{tabular}
\label{eClass}
\end{table*}
\twocolumn

\onecolumn
\begin{table*}[ht]
\centering
\caption{Visual inspection of 172 neighbour galaxies detected by eClass as ``late-type" with comparison
to Galaxy Zoo results. Out of the 275 companion galaxies that had Galaxy Zoo information available, 
172 galaxies were predicted to be late-type galaxies by the eClass function. The right column shows the results from the visual inspection and Galaxy Zoo-data of these eClass-defined late-type galaxies.}
\begin{tabular}{c c c c}
\multicolumn{4}{c}{eClass vs GalaxyZoo morphology classification} \\
\hline\hline
Galaxy type&eClassresults&Galaxy Zoo&Visual inspection \\ [0.2ex]
& (\#) & (\#) & (\#)\\
\hline
Total&172&172&172\\
Late-type&172&30&25\\
Early-type or indefinable&0&142&139\\ [0.2ex]
\hline
\end{tabular}
\label{GalaxyZoo1}
\end{table*}
\twocolumn

A visual morphology inspection by eye of the 183 eClass-defined late-type galaxies was also done (see Table 1).
Out of these only 70 were confirmed as late-type, while the rest appeared either undefinable or as early-types. 70 late-types 
confirmed out of 183 eClass-defined galaxies is a success rate of 38\%. Clearly using eClass as a predictor
of morphology was a failure for this data set.

A second morphology test using morphologies from the Galaxy Zoo project \citep{Lintott,Lintott2010}
was compared to the eClass results (see Table 2). For the 305 companion galaxies
275 had been classified in the Galaxy Zoo project\footnote{The SDSS casjobs interface allows one to obtain the
Galaxy Zoo classification for any object in their catalog with a corresponding SDSS Object ID.}.
From this sample of 275 with Galaxy Zoo morphologies, eClass yielded 172 late-type galaxies and 27 
early-type. For the 172 eClass defined late-type galaxies 30 were classified by Galaxy Zoo as ``Spirals",
30 as ``Ellipticals" and 112 as ``Unidentified."

Of those 30 galaxies selected by the Galaxy Zoo as late-type galaxies, 22 were identified as late-types by visual
eye inspection, while the remaining 8 were not possible to classify. Of the 30 galaxies selected by the Galaxy Zoo as
early-type galaxies, 28 were classified by eye as early-type galaxies and two could not be classified.
This demonstrates that Galaxy Zoo provides a relatively good estimate of morphology, especially in comparison with eClass. The clear advantage of the method is likely a result of the high internal consistency in the Galaxy Zoo morphological types.

\subsection{Colors of non-AGN galaxies and quasars}\label{sec:3.2}

Rest-frame colors of the stellar populations are easily calculated by applying Galactic extinction corrections 
and K-corrections to the five observed SDSS filters (u,g,r,i,z). In our study we used u-r color which is supposed to have low noise and is the most reliable indicator of the bimodality of the color populations \citep{Strateva2001}.

The Galactic extinction corrected (``dered'') magnitudes were taken from the SDSS casjobs database for both 
quasars and galaxies. For the galaxies, the K-correction could also be found in the SDSS casjobs database. 
The K-correction for our quasars were calculated assuming a universal power-law spectral 
energy distribution (SED) with the optical flux given by $f_{\nu}=\nu^{\alpha}$ and mean optical spectral index $\alpha$=-0.5. Since most of our quasars are at low redshift the K-correction is very small and the spectral index has changed little in this range \citep{Kennefick2008}. The K-correction for the quasar sample can be calculated as

\begin{equation}
K(z)=-2.5 \alpha\  log(1+z) - 2.5 log(1+z)
\end{equation}\label{eq1}

The absolute magnitudes of both galaxies and quasars were calculated via

\begin{equation}
M_{abs}=m_{obs} + 5 - 5log(D_L) - K(z)     
\end{equation}\label{eq2}

where D$_{L}$ is the luminosity distance in parsecs.

\begin{figure*}
 \centering
\includegraphics[scale=.8]{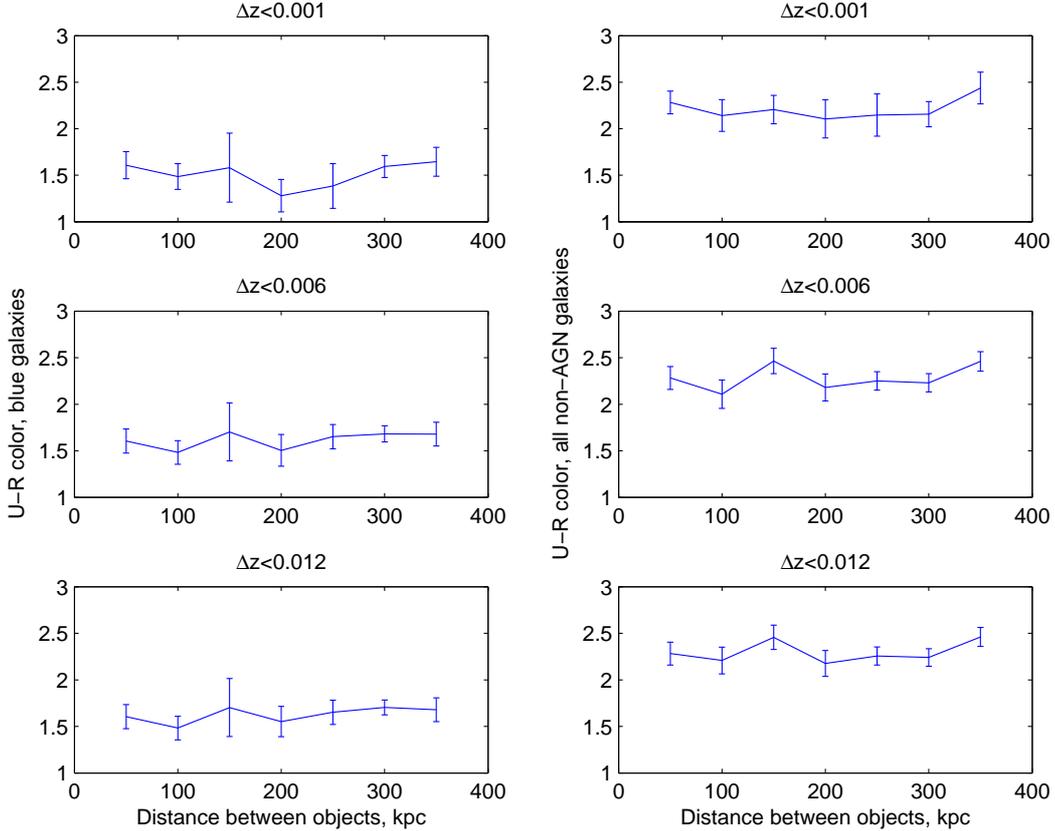}
\caption{$U_{e}-R_{e}$ color of galaxies versus distance to quasar. The left hand side plots
contain the $U_{e}-R_{e}$ colors of the bluer companion galaxies where $U_{e}-R_{e}<2.2$.
The right hand side plots contain all the non-AGN galaxy companions. Three different redshift difference cuts $|\Delta z|$ are pictured for the pairs, top: $|\Delta z|$=0.001, middle: $|\Delta z|$=0.006, bottom: $|\Delta z|$=0.012.} \label{Color}
\end{figure*}

Figure \ref{Color} shows how the internal extinction corrected $U_{e}-R_{e}$
color\footnote{The suffix ``e'' means ``internal extinction corrected" in this context.}
for the companion galaxy changes as a function of the projected distance between the quasar
and the non-AGN galaxy companion. We have chosen to perform binning of the data, and calculate
the average value of the color in each bin since this simplifies the visualization of the results.
Little change in color as a function of projected distance is observed.

%%graph1, Fig3.eps

\subsection{Surface-densities of non-AGN companions}\label{sec:3.3}

\begin{figure*}
 \centering
   \includegraphics[scale=.8]{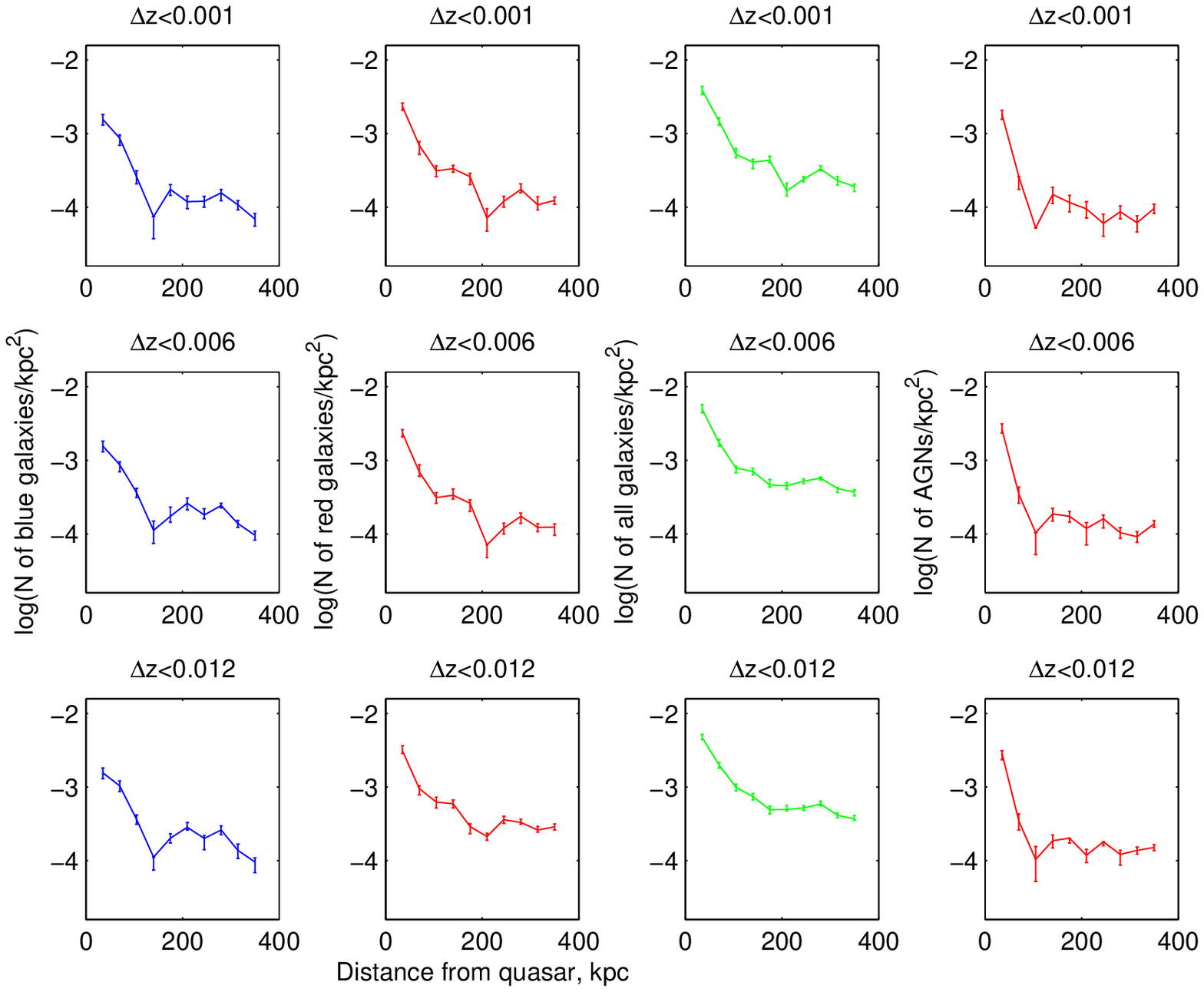}
\caption{Annular surface densities of different galaxies types around quasars. 
In the left column: blue galaxies with $U_{e}-R_{e}<2.2$, in the second column: red galaxies with $U_{e}-R_{e}>2.2$,
in the third column: all galaxies, in the most right column: all AGN neighbours. Three different redshift difference cuts $|\Delta z|$ are pictured for the pairs, top: $|\Delta z|$=0.001, middle: $|\Delta z|$=0.006, bottom: $|\Delta z|$=0.012.}
\label{endNFig85}%
\end{figure*}

The density of the environment around quasars may support a merger 
scenario if small over-densities around the quasars are observed.
To investigate this the surface densities of galaxies around quasars were calculated using

\begin{equation}
\rho=N_{distance}/((d_{1})^2-(d_{2})^2)\pi
\end{equation}\label{eq3}

where N is the number of galaxies in one bin (or ``annulus'') with outer radius $d_{1}$ and inner radius $d_{2}$.
The companion galaxies were divided into 10 bins a 35 kpc each in width. The error bars were generated via bootstrap resampling 100 times \citep[see][]{Tibby} and are plotted in figure \ref{endNFig85} with a 68.3\% confidence level, corresponding to 1 $\sigma$.

In general, the relative number of red background galaxies at distances further than d $>$ 100 kpc 
decreases with the smaller $|\Delta z|$ used. Perhaps this demonstrates the need to use small 
$|\Delta z|$ cuts when searching for galaxies that have projected separations of d $>$ 100 kpc. Indeed, 
previous studies \citep[e.g.][]{Yee1983} have shown that galaxies that lie within 100 kpc in projected 
distance have the largest chance to be a physical pair.

Figure \ref{endNFig85} shows large overdensites of galaxies around the quasars. The overdensity of AGN among the neighbours will be described in section \ref{sec:AGNsection}. This has been seen before,
but only in studies with large contaminations of background and foreground galaxies \citep[e.g.][]{Serber2006}. 
Ours is the first study with spectroscopic galaxies with well-defined redshifts made on such a small-scale environment near quasars (d $<$ 350 kpc) that yields similar results. However, the results presented herein 
are much more clear cut. This may give support to quasar formation via two scenarios.
The first involves the collapse of material into huge dark matter halos where smaller satellite galaxies
form around the massive quasar simultaneously. The second is a merger-driven scenario where massive objects get created via hierarchical assembling of smaller parts. In the next we will try to separate these two possibilities from each other. 

Another curious feature in the surface density plots is the minimum of blue 
galaxies in the bin around 150 kpc. The same feature is seen in the AGN surface density
(right most column of figure \ref{endNFig85}. A larger sample will be needed to 
determine whether this is a statistical artefact or something physical.

\subsection{Comparison to galaxy-galaxy pairs}

One way of identifying if the surface density enhancements are related directly to the nature of the quasars or
are present for all types of galaxies, is by comparing to the environment of field galaxies
of same luminosity range. We used a field galaxy sample derived from \cite{Way2011}
as described in Section \ref{sec:2.4}.

\begin{figure*}
 \centering
   \includegraphics[scale=.8]{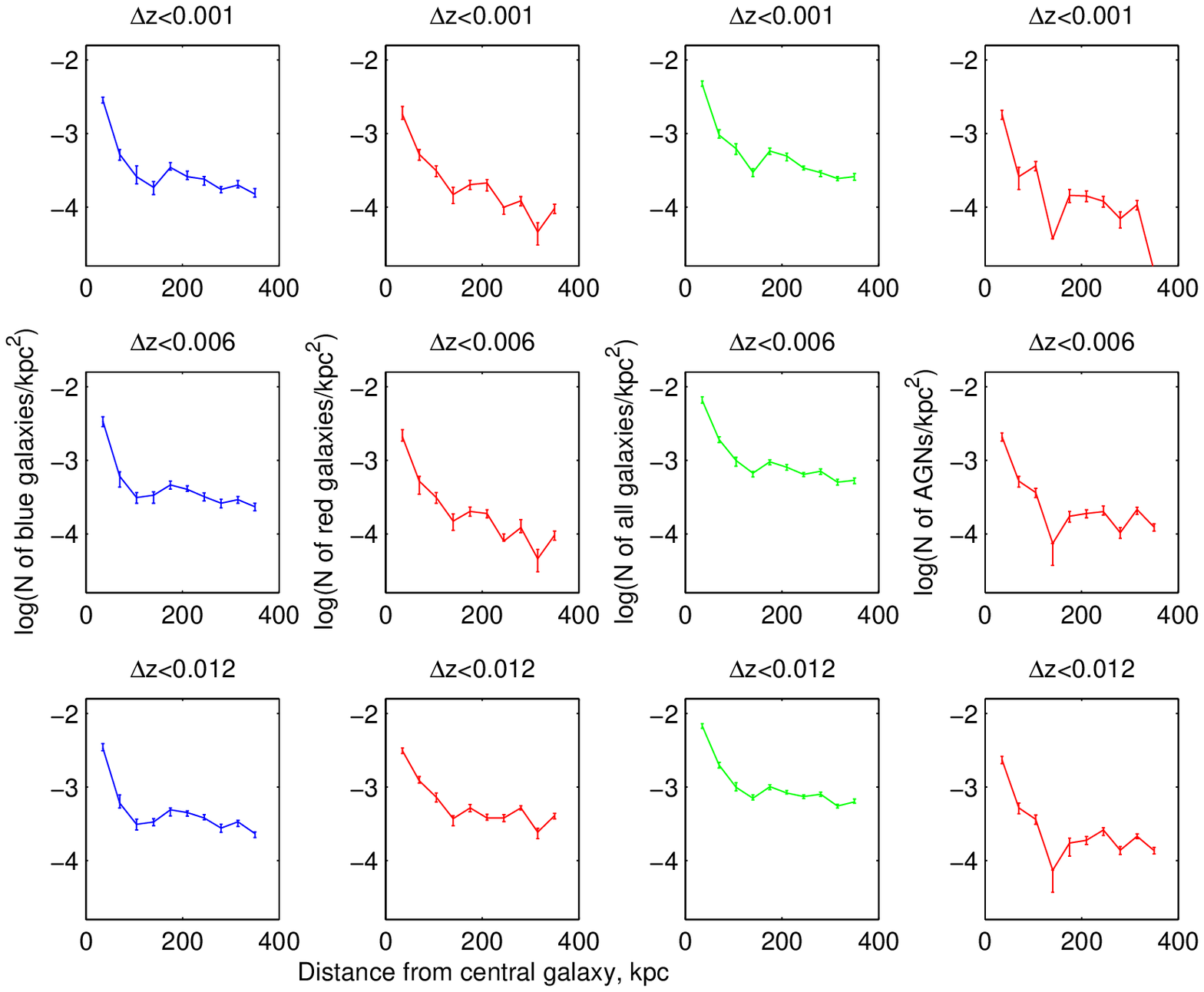}
   \caption{Annular surface densities of different types of galaxies around central field galaxies.
In the first column: blue galaxies with $U_{e}-R_{e}<2.2$, in second column: red galaxies with
$U_{e}-R_{e}>2.2$, in third column: all non-AGN galaxies, in the most right column: all AGN neighbours.
Three different redshift difference cuts $|\Delta z|$ are pictured for the pairs, top: $|\Delta z|$=0.001, middle: $|\Delta z|$=0.006, bottom: $|\Delta z|$=0.012.}
               \label{ggNFig85}%
     \end{figure*}

Figure \ref{ggNFig85} show surface density plots of the field galaxy
environment using normal (non-AGN) field galaxies with a similar absolute magnitude range to the
quasar sample. There is a certain skewness towards higher luminosities in the quasar hosts if 
comparing to the field galaxies. We have tried to restrict the problems of different luminosity distributions by selecting field galaxies within the same luminosity range as the quasars. 

\begin{figure*}
 \centering
   \includegraphics[scale=.8]{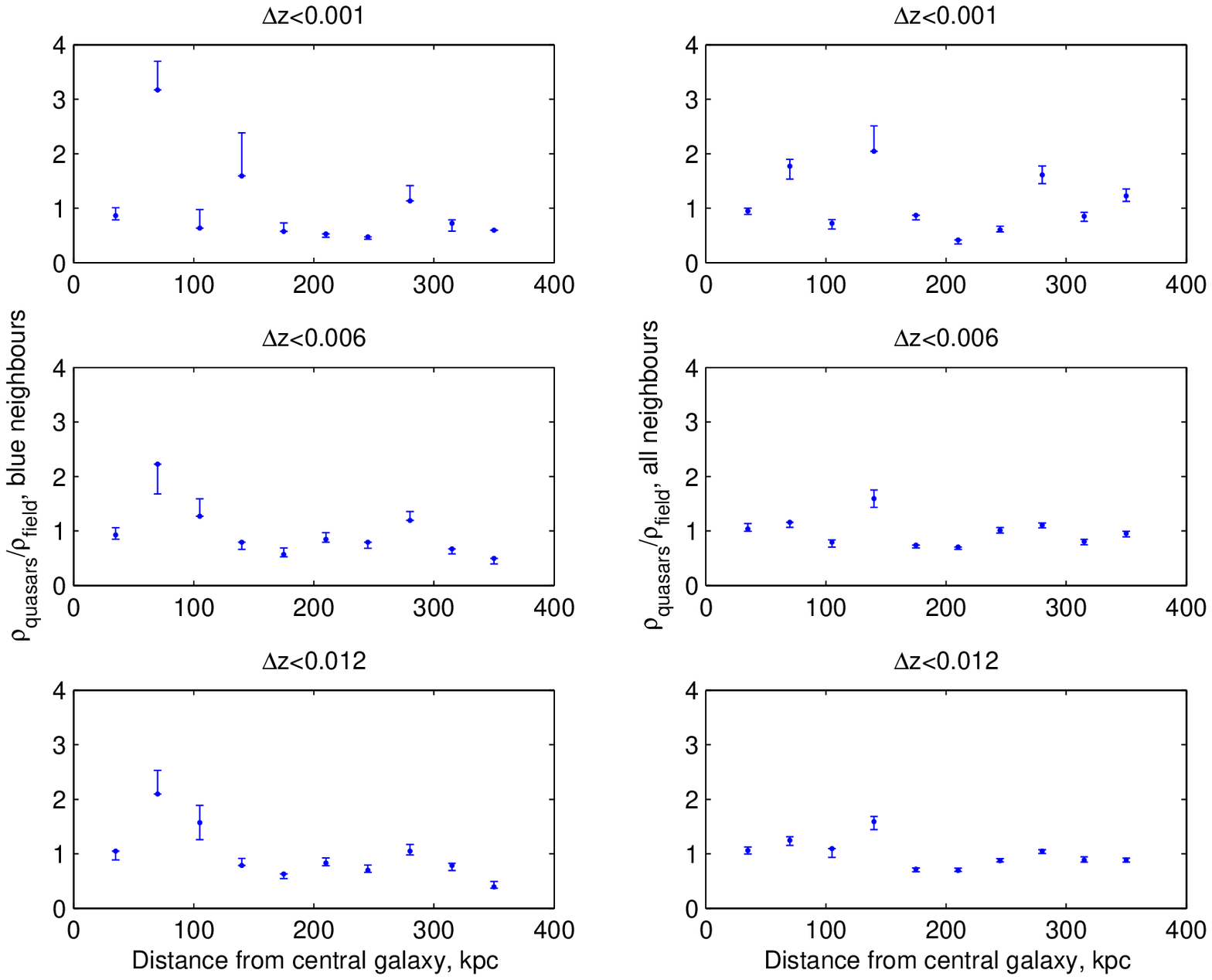}
   \caption{Results from the spectroscopic redshift samples. Ratio of annular surface densities of galaxies around quasars ($\rho_{quasars}$) and the annular surface densities of galaxies around central non-AGN field galaxies ($\rho_{field}$) as function of projected distance. The field galaxies were chosen to be inside the same magnitude range as the quasars in order to compare objects of same mass range, having -24 $ < M_{r} < $-19. The plot includes AGN among the companion galaxies. In the left column: blue galaxies with $U_{e}-R_{e}<2.2$, in right column: all galaxies. Three different redshift difference cuts $|\Delta z|$ are pictured for the pairs, top: $|\Delta z|$=0.001, middle: $|\Delta z|$=0.006, bottom: $|\Delta z|$=0.012. The ratio is normalized for the difference in sample size.}
     \label{kvot}%
     
\end{figure*}

We now would like to know if the increase in the surface density of field satellites is as steep as the one of neighbour galaxies to quasars. Any increase at short projected separations could mean that the overdensity around quasars is steeper than around field galaxies and forms in different process. Figure \ref{kvot} shows the ratio of surface densities for satellites of quasars versus the field sample. For the sake of the sample sizes we included AGN among the neighbours in this calculation. We see an increase of the ratio at short projected separations. A potential issue here is the difference in the redshift distribution between the two samples. The field galaxy-galaxy pairs range between 0.03 $<$ z $<$ 0.1 while the quasar-galaxy pairs range 0.03 $<$ z $<$ 0.2. We tried restricting our study by using volume-limited subsamples of quasar-neighbour and field-neighbours within this redshift range 0.03 $<$ z $<$ 0.1, together with a cut on the luminosity magnitude on central field galaxies and quasars on $M_{r}$ $>$ -24. The too small sample size made only the largest cut ($|\Delta z|$) any interesting, confirming a steeper increase in the ratio of blue neighbours.

\begin{figure}
 \centering
   \includegraphics[width=8cm]{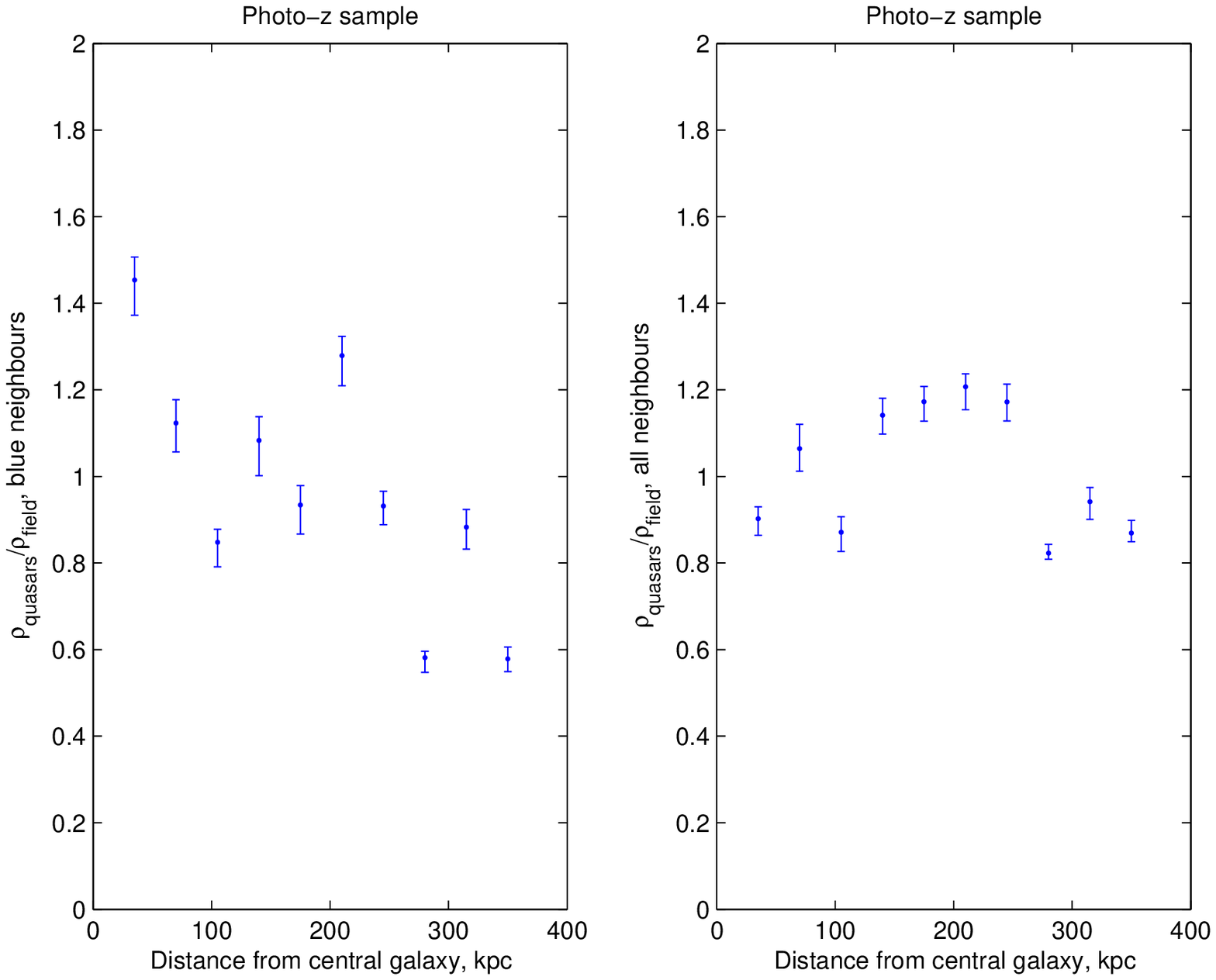}
   \caption{Results from the photometric redshift samples. Ratio of annular surface densities of galaxies around quasars ($\rho_{quasars}$) and the annular surface densities of galaxies around central non-AGN field galaxies ($\rho_{field}$) as function of projected distance. The field galaxies were chosen to be inside the same magnitude range as the quasars in order to compare objects of same mass range, having -24 $ < M_{r} < $-19. The quasar-galaxy pairs were selected to be inside the same redshift range as the field galaxy pairs, 0.03 $<$z $<$ 0.1. The plot includes AGN among the companion galaxies. In the left column: blue galaxies with $U_{e}-R_{e}<2.2$, in the right column: all galaxies. $|\Delta z|$ $<$ 0.03. The ratio is normalized for the difference in sample size.}
     \label{PhotoRatio}%
     \end{figure}

The fiber collision constraint could eventually lead to an underestimation of the clustering of blue neighbours near quasars at short projected separations. We therefore do a similar plot with the two photometric redshift samples of quasar-galaxy and field-galaxy pairs where we also select the objects within the same absolute magnitude ranges and within the field galaxy redshift range 0.03 $<$ z $<$ 0.1 for a direct comparison. In total, 830 quasar-galaxy pairs and 2907 field galaxy-galaxy were used in this plot. Figure \ref{PhotoRatio} confirms that the clumping of blue neighbours might be stronger than observed with the spectroscopic neighbours. The trend significance in the two different plots (figure \ref{kvot} and figure
\ref{PhotoRatio}) is the same ($\sim$ 3.5$\sigma$), but the photometric redshift samples
show a 20-25 \% steeper clustering of blue galaxies around quasars than does the spectroscopic redshift sample.

Also the effective r$\sim$ 19 limit for spectroscopic targeting biases us to miss pairs with a low-luminous neighbour at higher redshift. While the magnitude limit would not cause any problems in case there would be no correlation between companion galaxy luminosity and the distance from quasar, and neither is too problematic for the case of an increased amount of massive galaxies near the quasar, it could on the other hand create difficulty in detecting an increased amount of less massive galaxies at short projected separations. A more tolerant magnitude limit of spectroscopic targeting could permit us to detect less luminous companions. We therefore suspect that the clustering of low-luminous, blue galaxies around quasars might be even higher than observed at the moment.

At the end of the day, this revisits an old issue about how the occurrence and luminosity of AGN relates to the properties of the host galaxies-- an issue we thought was clear long ago, but apparently is not so clear.

\subsection{AGN fraction in neighbours\label{sec:AGNsection}}

\begin{figure}
 \centering
   \includegraphics[width=8cm]{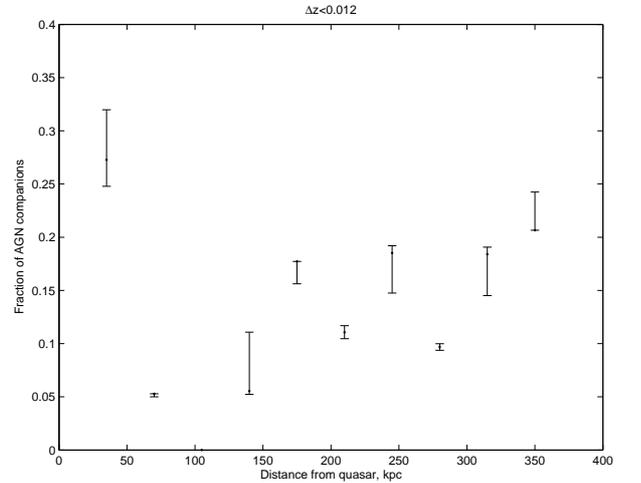}
     \caption{The fraction of AGN companions as a function of distance from quasars. Only the largest
redshift difference cut $|\Delta z|$=0.012 is used. In total 69 quasar-AGN pairs were detected, 
where 14 reside in the closest bin.}
               \label{AGNfraction}%
     \end{figure}

We have a small sample consisting of only 69 quasar-AGN pairs, but we are still interested
to see if we can get any rough indication of how the AGN fraction in those might vary
as a function of distance. Figure \ref{AGNfraction} represents how the AGN fraction in the sample 
changes as a function of distance from the quasars. In the figure, the AGN fraction in annuli 
drops off steadily until 100-150 kpc, where steeply increases again. Out of the quasar-AGN pairs,
14 reside in the closest bin. However, the rough estimation might be just random features unless the same can be shown with a much larger sample of quasars with neighbours.

A similar plot for the field-galaxy pairs revealed no changes in the AGN fraction as function of distance. One might speculate that there could exist two different processes influencing the occurrence of AGN in companion galaxies. While one of these processes would be more important at low galaxy densities, the other one could be directly associated with mergers or interactions and therefore only is observed in the closest bin.

\subsection{Dust in non-AGN neighbour galaxies}\label{sec:3.4}

\begin{figure}
 \centering
   \includegraphics[width=8cm]{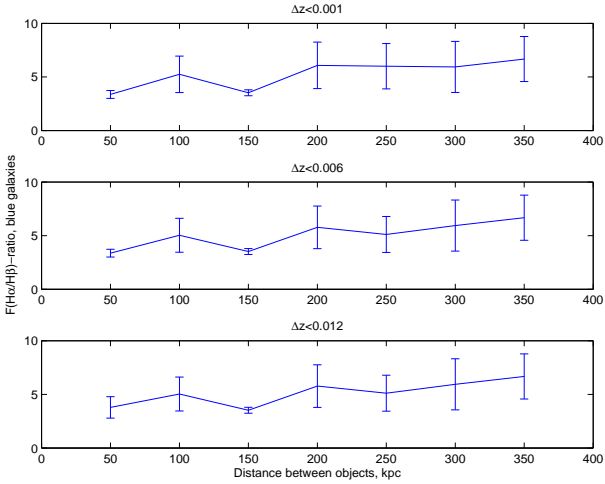}
     \caption{Dust extinction: the flux ratio F(H$\alpha$/H$\beta$ ) in blue galaxies with $U_{e}-R_{e}<2.2$.
Three different redshift difference cuts $|\Delta z|$ are pictured for the pairs, top: $|\Delta z|$=0.001, middle: $|\Delta z|$=0.006, bottom: $|\Delta z|$=0.012. The individual errors are unweighted in the
calculations of mean values and error bars. Only objects with emission line
fluxes that are greater than three times the measured error are included.}
               \label{Figure9}%
     \end{figure}
     
We are also interested in seeing how the dust content in the surrounding
galaxies varies as a function of distance from the quasar. One standard way of measuring the dust
content is by using the H$\alpha$/H$\beta$ emission line flux ratio. Consequently, we
can only use galaxies with fluxes $f >0$ in H$\alpha$ and H$\beta$.
 
Figure \ref{Figure9} shows no significant correlation of F(H$\alpha$/H$\beta$) with distance from the quasar.
Unfortunately, not many of our galaxies have H$\beta$ line.

\subsection{Star-formation rate (SFR)}\label{sec:3.5}

By the mid-1970's several studies \citep[e.g.][]{Bushouse1987} suggested that tidal interactions and shocks
could cause increased star-formation in galaxies. During mergers large galaxy disks could be disrupted by tidal effects, causing 
large gas flows and thus fueling star formation in the central regions \citep[e.g.][]{Barnes1996}. 
More recent simulations \citep{Martig2008} have suggested that in mergers between two galaxies 
occuring near a large tidal field (e.g. near a group of galaxies) will increase the star-formation rate 
in the merger. Recently, many groups have also observed an increased star-formation rate during mergers 
\citep[e.g.][]{Nikolic} and also in the neighbourhoods of AGN (e.g. CW06). 
	We calculate the star formation rate using the H$\alpha\ \lambda$6563 emission line. 
In star-forming galaxies the H$\alpha$ emission originates from photoionization by massive, 
short-lived stars. The advantage of the H$\alpha$ line is that it 
suffers least from extinction (in the optical) and depends least on the metallicity
of the ionized gas \citep[e.g.][]{Moustakas2006}.

Using the SDSS measured H$\alpha$ emission line the \cite{Bergvall} calibration
was utilized to estimate the star formation rate of our galaxy sample:

\begin{equation}\label{eq4}
L(H\alpha)= SFR \times 1.51*10^{34}  [W]
\end{equation}
and

\begin{equation}\label{eq5}
L(H\alpha)=4 \pi D_L^{2} \sqrt{2\pi} \sigma h 10^{-20}
\end{equation}

where SFR is the star formation rate in solar masses per year, $\sigma$ and $h$ are the width and 
height of the H$\alpha$ emission line, $D_L$ is the luminosity distance in Mpc and the emission line
luminosity is expressed in Watts.

As an alternative way of measuring the star formation rate we used the [\ion{O}{ii}]- emission line. 
In SDSS related studies [\ion{O}{ii}] $\lambda$3727 is mainly useful when trying to calculate the star formation rate
in galaxies at redshifts higher than z=0.4 where the H$\alpha$ emission line is redshifted
beyond the red limit of the SDSS spectrograph. The [\ion{O}{ii}]-emission line is more affected by the metallicity
and internal dust extinction in galaxies. To reduce the reddening and metallicity effects,
we use the K98 calibration \citep{Kewley2004} for calculating the SFR from the [\ion{O}{ii}]- emission line:

\begin{equation}\label{Kewley}
SFR=(1.4 \pm 0.4) \times 10^{-41} L([\ion{O}{ii}])
\end{equation}

where the L([\ion{O}{ii}]) is given in erg s$^{-1}$.

\begin{figure*}
 \centering
  \sidecaption
   \includegraphics[width=12cm]{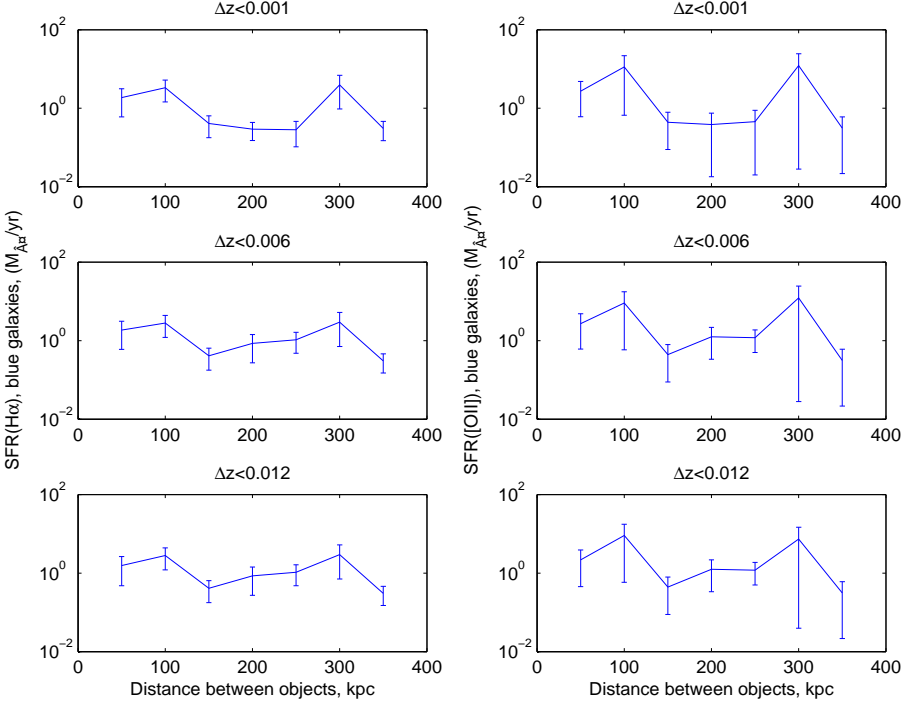}
     \caption{Star formation rate in blue companion galaxies in solar masses per year. Left column shows SFR from from H$\alpha$-flux. Right column shows SFR calculated from [\ion{O}{ii}]-flux. Three different redshift difference cuts are pictured for the pairs, top: $|\Delta z|$=0.001, middle: $|\Delta z|$=0.006, bottom: $|\Delta z|$=0.012. Galaxies with SFR$>$100 and individual errors in SFR$_{error}$ $>$1000 are excluded. The individual errors are unweighted in the calculations of mean values and error bars.
Only objects with emission line fluxes that are greater than three times the error in the emission line flux are included.}
     \label{SFR}%
\end{figure*}

Figure \ref{SFR} shows the average star-formation rate in blue companion galaxies calculated from H$\alpha$
flux versus distance from the quasars (left column), as well as SFR calculated from [\ion{O}{ii}] (right column).
According to the figure \ref{SFR} SFR via H$\alpha$ remains unaffected as a function of distance,
while star formation rate from the [\ion{O}{ii}]-line increases in proximity to the quasar.
The cause of this inconsistency could arise from variations in metallicity or dust extinction as a function of distance
from the quasar.  Another possibility may be that a number of AGN galaxies were not properly identified
in the BPT diagram and therefore create a false trend in the [\ion{O}{ii}] derived SFR.

\begin{figure*}
 \centering
  \sidecaption
   \includegraphics[width=12cm]{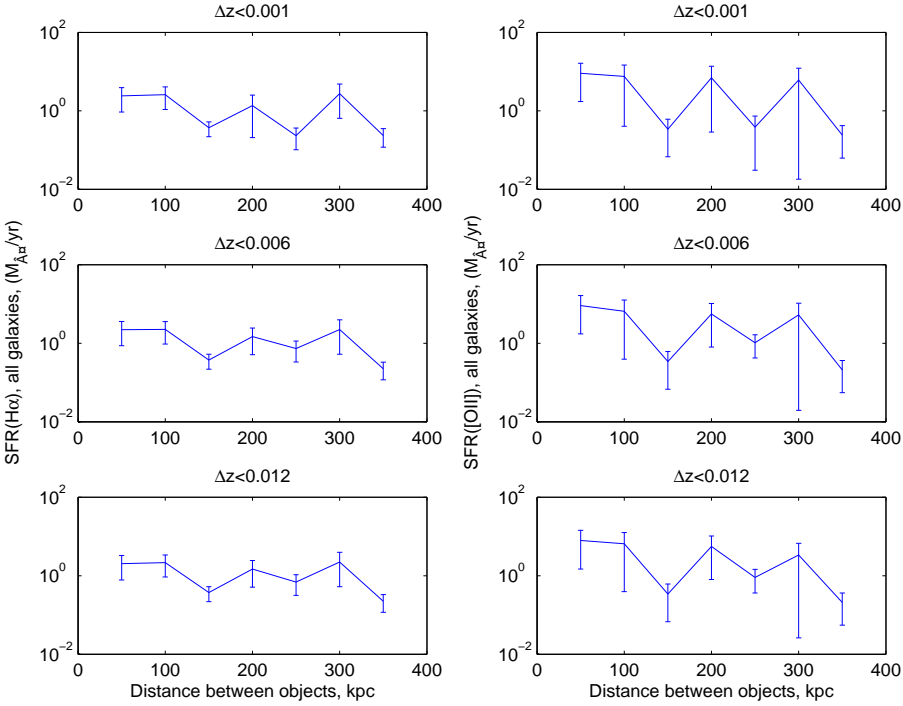}
     \caption{Star formation rate in all non-AGN companion galaxies in solar masses per year. Left column shows SFR from from H$\alpha$-flux. Right column shows SFR calculated from [\ion{O}{ii}]-flux. Three different redshift difference cuts are pictured for the pairs, top: $|\Delta z|$=0.001, middle: $|\Delta z|$=0.006, bottom: $|\Delta z|$=0.012. Galaxies with SFR$>$100 and individual errors in SFR$_{error}$ $>$1000 are excluded. The individual errors are unweighted in the calculations of mean values and error bars. Only objects with used emission line fluxes that are greater than three times the error in emission line flux are included.}
     \label{SFRall}%
     \end{figure*}

Figure \ref{SFRall} shows the average star-formation rate in all non-AGN companion galaxies with detected H$\alpha$ and [\ion{O}{ii}].
The SFRs from [\ion{O}{ii}] and H$\alpha$ both decrease as a function of 
distance from the quasar, but the correlation in the SFR from [\ion{O}{ii}] is still clearer.

\subsection{Degree of ionized gas in non-AGN neighbour galaxies}\label{sec:3.6}

The flux ratios F([\ion{O}{iii}]/[\ion{O}{ii}]) and F([\ion{O}{iii}]/H$\beta$) are useful since they provide information on the ionization state of the gas in the galaxies. High values could occur both in the presence of an AGN, in areas with high star-formation activity and from shock-heating of the gas.

The flux ratios are sensitive to metallicity and low metallicities decreases the cooling efficiency which, in turn, increases the electron temperature and give higher ratios \citep{Searle1971}. The ratios may be further modified by shocks, whose presence also will increase the ratios \citep[e.g.][]{Raymond,SuthDop}.

\begin{figure*}
 \centering
  \sidecaption
  \includegraphics[width=12cm]{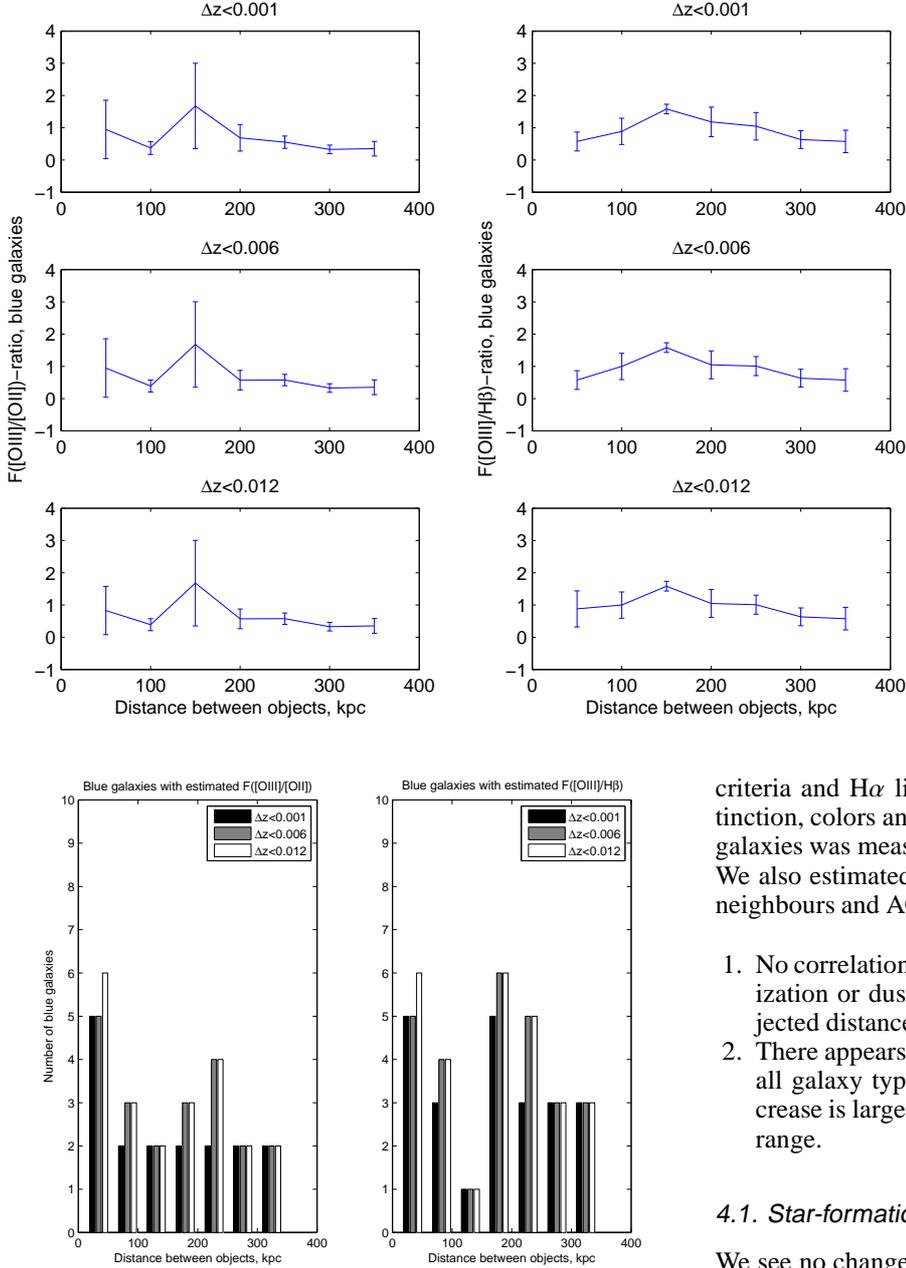}
     \caption{Ionization in blue companion galaxies. Left column shows the flux ratio
F([\ion{O}{iii}]/[\ion{O}{ii}]). Right column shows the flux ratio F([\ion{O}{iii}]/H$\beta$). 
Three different redshift difference cuts are pictured for the pairs, top: $|\Delta z|$=0.001, middle: $|\Delta z|$=0.006, bottom: $|\Delta z|$=0.012. Galaxies with ionization$>$7
are not included since they are believed to be AGN that have contaminated the sample despite using
the BPT criteria. Galaxies with individual errors$>$1000 in ionization are excluded.
The individual errors are unweighted in the calculations of mean values and error.
Only objects with emission line fluxes greater than three times the error in emission line flux are utilized.}
     \label{slutIonization}
     \end{figure*}

\begin{figure}[htb!]
 \centering
   \includegraphics[width=8cm]{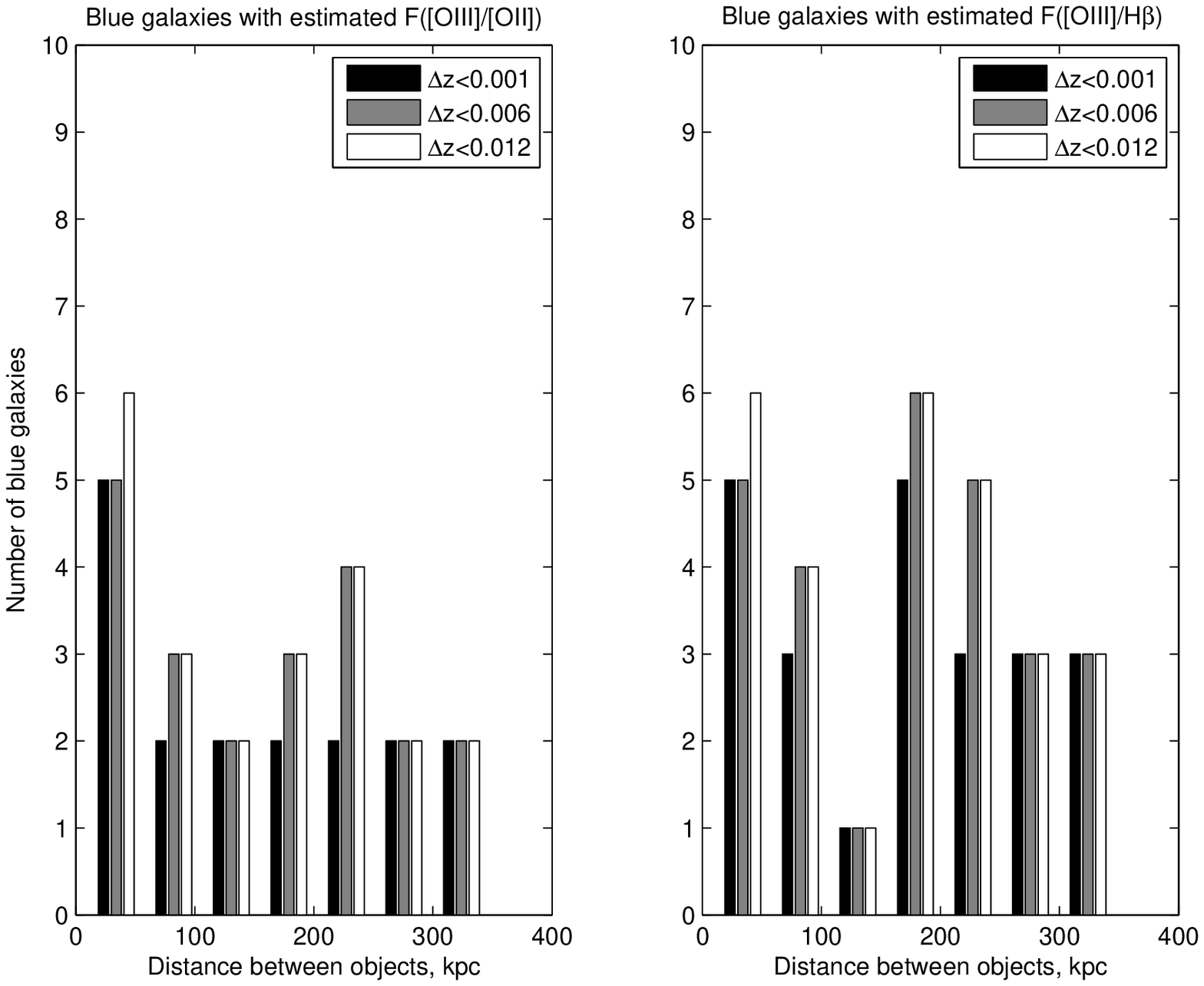}
     \caption{Left column: number of galaxies with [\ion{O}{iii}] and [\ion{O}{ii}] measured.  
Right column: number of blue galaxies with [\ion{O}{iii}] and H$\beta$ measured.
The three $|\Delta z|$ are represented by different bar colors. Galaxies with ionization$>$7 are
not included since they are believed to be AGN that have contaminated the sample despite using
the BPT criteria. Galaxies with individual errors$>$1000 in ionization are excluded.
The individual errors are unweighted in the calculations of mean values and error.
Only objects with emission line fluxes greater than three times the error in emission line flux are utilized.}
     \label{NFig75}
     \end{figure}

In figure \ref{slutIonization} the F([\ion{O}{iii}]/[\ion{O}{ii}])-parameter shows little change in ionization for blue galaxy companions as a function of distance from the quasar. The same holds true for the F([\ion{O}{ii}]/H$\beta$)-ratio shown in the same figure. In figure \ref{NFig75} the number of galaxies with measurable ionization is plotted. The fact that the ionization cannot be measured for the entire sample makes it difficult to drawn any substantive conclusions. We cannot observe any correlation of degree of ionization of the gas in the neighbour galaxies as a function of distance to the quasars. Clearly we suffer the fact that only a small fraction of the sample could be examined due to the lack of line data. This hinders us to draw any firm conclusions.

\section{Discussion and Conclusions}\label{sec:4}

Using a number of derived quantities from the photometry and spectroscopy of the SDSS DR7
we have investigated the properties of 305 quasar-galaxy associations at redshifts of
0.03 $<$ z $<$ 0.2 and projected distances less than 350 kpc,

Three different methods for morphology classification of the quasar companion galaxies
were utilized: eClass, Galaxy Zoo and by eye. Galaxy color and spectral line measurements 
were corrected for internal dust extinction in the non-AGN companion galaxies. AGN were 
separated from the other neighbour galaxies using BPT-diagram combined with \cite{Kauffmann2003} 
criteria and H$\alpha$ line width. The degree of ionization, dust extinction, colors and star-formation rates (SFR) 
for the non-AGN galaxies was measured as function of distance from the quasars. We also estimated 
the surface density distribution of non-AGN neighbours and AGN neighbours. Some conclusions are:

\begin{enumerate}

\item No correlation was found between color, SFR, degree of ionization or dust extinction in the neighbour galaxy 
and projected distance from the quasar.

\item There appears to be a large increase in the surface density of all galaxy types closer 
than 100 kpc to the quasar. The increase is larger than for field galaxies of the same luminosity range.

\end{enumerate}

\subsection{Star-formation rates and colors}

We see no change in star-formation rate from H$\alpha$, color, degree of ionization or dust extinction in the companion galaxies as a function of projected distance from the quasars.

This is contrary to what several other studies have reported on quasar environments (e.g. CW06) where change in colors and SFR in close neighbours have been reported
at short projected separations. Different studies could have differing SFR conclusions because the methods used to separate AGN from normal galaxies vary between the studies.

The general lack of correlations in the quasar non-AGN companions as function of distance from the quasar could be a result from having a sample consisting mainly of faint quasars. If one wishes to see effects from the hypothesized AGN feedback on the distance scale up to 350 kpc, one might need to perform a study with a sample consisting of very strong high-redshift quasars. Since the present day catalogues of quasars have too few observable companion galaxies we cannot do a similar high-redshift study yet. At the moment the SDSS DR7 does not allow either study due to the far too small sample size one obtains from the present sample and the lack of objects with spectroscopic redshifts above z $>$ 0.2. 

The lack of observable effects on companions can also be a consequence of quasar influence being episodic rather than continuous. A short episode of effects from the quasar on the companion could perhaps not be disentangled by the color, star-formation and ionization tests unless the gas would be completely swept away. Ionization from the quasar on the interstellar medium in the companion galaxy might regenerate to a normal state with time and a past burst of star-formation in the nearest neighbours could already have died out, leaving no trace in our observations. Therefore it might be more advantageous for future studies to concentrate on a combination of total masses of the atomic and molecular gas content together with spectral history modeling for the nearest neighbours to disentangle the possibilities of a past episode of starburst or quenching.

We could still expect to see effects of close interactions on the SFR or color as many other studies \citep[e.g.][]{Kennicutt, Ellison, Nikolic} reported observations that interactions and mergers can enhance the SFR. However there are contraindications that interactions are not enough for causing starbursts unless the galaxies actually are merging, and that even mergers might not be a sufficient condition \citep[e.g.][]{Bergvall2003}. This might be consistent with that we cannot see any increase of SFR or change in color in our quasar neighbours at short projected separations.

\subsection{AGN fraction of neighbours}

Our results suggest that there might be a drop of the annular AGN fraction 
among the neighbours to quasars, and a steep increase the closest 50-100 kpc.

A larger study of AGN pairs by Liu et al. (2011) shows steadily dropping cumulative AGN 
fraction represents a large sample with 1286 AGN-AGN pairs. Their pairs are a mixture of 
all kind of active galaxies with a S/N ratio $>$ 3 for each emission-line used in the AGN 
identification; narrow-line galaxies derived from the MPA-JHU SDSS DR7 catalogue \citep{Aihara2011}, 
narrow-line quasars \citep{Reyes2008}, broad-line galaxies from \citep{Hao2005} and 
broad-line quasars from the quasar catalogue \citep{Schneider2010}. In Liu's study they show 
that for AGN-pairs with emission lines having a S/N ratio$>$15, 80\% of the pairs display 
tidal features, something that supports merging and tidal interactions behind AGN activity. 

The implicit assumption for the selection of Liu's parent AGN sample is that all AGN are 
the same type of object observed with different viewing angle. If there are clearly different 
formation processes for different types of active galaxies and the quasars in our sample are not formed in 
the same way as the AGN in their sample, we would see a different behavior of the AGN fraction 
versus projected distance between the objects in comparison to our study with quasar-AGN pairs.

We morphologically examined the 14 quasar-AGN pairs within 50 kpc, where 8 showed tidal features. 
An interesting question is whether our they are real AGN or possibly normal galaxies contaminated by light from the quasar. It is possible that if using a higher S/N-cut could decrease the fraction in the nearest bin, where the contamination from the quasars also has the highest probability to influence our emission lines. 
Maybe it could also be explained in the light echo from quasar such that has been demonstrated in 
Hanny's Voorwerp \citep{Lintott2009}. In such case the contamination would be separated largely by the 
time of light travel between the two quasar and the neighbour, in such a way that the quasar already 
would have evolved. We note that these 14 neighbours have rather low F([\ion{O}{iii}]/H$\beta$) $\sim$ 1. X-ray measurements indicative of non-thermal source could be a good way to find out if they are real AGN 
or not.

\subsection{Enhanced surface densities}

We plotted the surface densities of the blue neighbour galaxies, the red galaxies and all galaxies combined and found a large increase in surface density of all galaxy types at bins closer than 150 kpc to the quasar. Especially high is the increase in blue gas-rich galaxies and AGN if one compares to the background of galaxies further away from the quasar (d $>$ 150 kpc). In case mergers are responsible for quasar activity, one could imagine that the
nearby blue galaxies and AGN also are products of local mergers. But this hypothesis would require modeling of star formation history of the objects, which is beyond the scope of this paper.

At the same time an increased density around the quasars does not necessarily imply a merger scenario, even though the temptation to assign this explanation to the result might be strong. Monolithic collapse models can predict the formation of smaller objects around more massive ones when both smaller and larger objects collapse in the gravitational well of the same halo \citep[e.g.][]{Press}. The $\Lambda$CDM model itself predicts a large number of satellite galaxies around massive galaxies, something that is still difficult to know whether it is the case or not here. 

Our comparison to field galaxies of the same luminosity distribution as the quasars shows that at short projected 
separations quasars might have more satellites than the field galaxies of similar luminosity distribution, and 
particularly blue satellites. This is possible to interpret as that quasars form from mergers 
of gas-rich (blue) objects and do not form in the same way as field galaxies. 

However, the most interesting feature of the data might lie in the difference between the total
number of non-AGN neighbours to quasars or field galaxies in the $|\Delta z|$=0.001 and 
$|\Delta z|$=0.012 samples, see table 3. As indicated from the table, the fraction of satellites
residing in the smaller subsample is larger around quasars than around field galaxies, 
which is another support for stronger clustering of satellites around the quasars.

Also CW06 found support in their studies of the surface density distributions of galaxies close to clusters, 
close to field galaxies and close to quasars that galaxies in the neighbourhood of field galaxies are found in 
lower density environment than galaxies around quasars. Similar results have been obtained by other groups
\citep[e.g.][]{Serber2006, Strand2008}.

\begin{table*}[ht]
\caption{Number of non-AGN satellites around quasars and field galaxies within 350 kpc. The number
of satellites in the subsample $|\Delta z|$=0.001 (column 2) is compared to the main sample $|\Delta z|$=0.012 (column 3). The fraction of satellites in the subsample $|\Delta z|$=0.001 (column 4).}
\centering
\begin{tabular}{c c c c}
\multicolumn{4}{c}{Clustering of neighbours} \\
\hline\hline
Central galaxy & $|\Delta z|$=0.001 & $|\Delta z|$=0.012 & fraction\\ [0.2ex]
     & (\#) & (\#) & (dimensionless)\\
\hline
Quasars & 138 & 235 & 59\% \\
Field & 161 & 326 & 49 \% \\ [0.2ex]
\hline
\end{tabular}
\label{Sample}
\end{table*}

How could we further test whether the increased number of satellites around quasars is a product of 
hierarchical structure formation or of monolithic collapse? In a merger scenario, the observed neighbour galaxies were not involved in forming our quasars, and we see only those satellites that are left after the process. One way would be to construct a SDSS sample of field quasars at low-redshifts (0.03 $<$ z $<$ 0.2) but that have no companion closer than 350 kpc. In case of merger scenario being the secret of quasar activity, these quasars should have an average luminosity that is fainter than in the sample with quasars that have companion galaxies. A second way would be to use the local velocity dispersions of neighbours as indicator of tendency to merging. However, this would demand much larger samples as the lack of accuracy in SDSS redshifts might make it difficult to draw any conclusions otherwise.

An additional way is to approach the problem is by using modelling to predict the number of lower-mass objects around quasars of the average luminosity in our quasar sample \citep[e.g.][]{Carlberg1990}. Similar attempts of investigating the clustering of quasars have been done \citep[e.g.][]{HopkinsCluster} where an excess of clustering could be shown, but where the results could not be clearly put in context of any given structure formation scenario. 

\subsection{Missing AGN and blue galaxies?}

In this section we would like to offer the reader some more speculative statements on the data 
aiming to nurture a curious mind, yet without the intention of being conclusive or complete.

\begin{figure*}
 \centering  
\sidecaption
   \includegraphics[width=12cm]{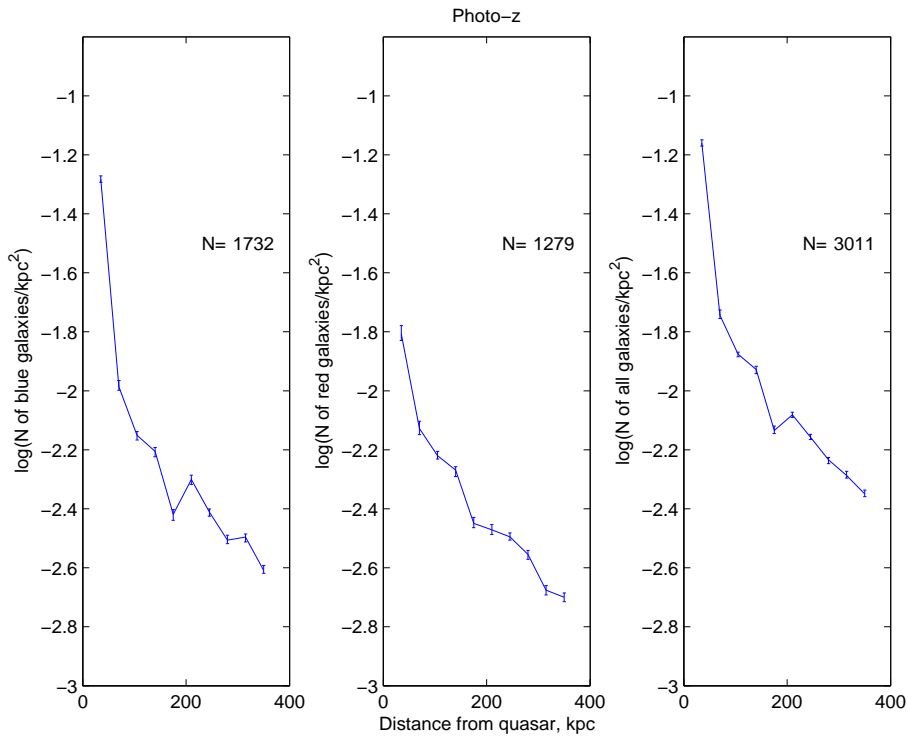}
   \caption{Results from the photometric redshift samples. Annular surface densities of neighbour galaxies around quasars. In the first column: all blue neighbours with $U_{e}-R_{e}<2.2$, in middle column: all red neighbours with $U_{e}-R_{e}>2.2$, in the most right column: all neighbours. The total number of galaxies is indicated in each plot. Only one redshift difference cut $|\Delta z|$ $<$ 0.03 is used. AGN are included among the neighbours.}
               \label{PhotoKvasarNFig85}%
     \end{figure*}

A curious thing about the surface density plots is the minimum of blue galaxies and AGN in the bins around $\sim$ 150 kpc on a statistical significance level of $\sim$ 1.5 $\sigma$. This could be a product of noisy data or low number statistics, but at the same time we see a slight increase of red galaxies in the 35 kpc closer surface density bin. We did similar surface density plots for the photometric pair samples with only color separation (blue and red) among the neighbours, see figure \ref{PhotoKvasarNFig85}. For the quasar-galaxy pairs with a blue neighbour (1732 pairs), the 150 kpc gap again could be spotted on a $\sim$ 2 $\sigma$ significance level. Nothing similar was spotted for the photometric field galaxy pairs with blue companions.

Could this mean that those missing blue galaxies got trapped by the gravitational field of the quasar and got their gas stripped off (hence became red) as they fell into circular orbits around the quasar or lost it due
to ionization pressure from the quasars?

It should be noted that the surface densities are shown on a logarithmic scale. It may seem like the loss of blue galaxies and AGN neighbours in the $\sim$ 150 kpc bin is larger than the gain of red galaxies. Perhaps some galaxies lost the gas when falling into bound orbits. Or maybe there is an unseen population of very faint galaxies 
in this bin. 

What if the galaxies simply not had a chance even to form any stars in their recent
star formation history, before getting trapped by the quasars? A study of Milky Way satellites \citep{Grcevich}
has revealed that within 270 kpc of the Milky Way, almost all satellite galaxies have lost their gas
and are spheroidal. This is support for tidal stripping processes within the distance 
range in our own study. What if something similar has happened to our the blue non-AGN companions at $\sim$ 150 kpc? 
These highly speculative scenarios would however need careful measurements of the velocity 
fields of the close quasar companion galaxies in order to follow how the dynamics of the gaseous content
of the companion galaxies changes around the minimum.

Many thoughts remain as pure speculations until we find a way to increase our 
neighbour galaxy sample without losing precision in redshift estimates. It is difficult to know whether
the minimum is real or simply an artefact from having too few quasar-galaxy pairs. Further studies with and increased number of galaxies will be required to confirm (or reject) this detection.

%Many thoughts remain as pure speculations until we find a way to increase our 
%neighbour galaxy sample without losing precision in redshift estimates. It is difficult to know whether
%the gap is real or simply an artefact from having too few quasar-galaxy pairs.

%\begin{acknowledgements}
\acknowledgements
      The author wishes to thank Michael J. Way and Nils Bergvall for many good advice on the draft and help during the project. Thanks to William C. Keel for very constructive comments on my manuscript that significantly helped to improve it and to Kjell Lundgren, Martin Lopez-Corredoira, Lars Mattson, Allen Joel Anderson and G\"oran Henriksson for fruitful discussions on the results. Lastly, she wishes to thank Pianist for always being available for support and help.

This research was done with help of the SDSS. Funding for SDSS-II has been provided by the Alfred P. Sloan Foundation, the Participating Institutions, the National Science Foundation, the U.S. Department of Energy, the Japanese Monbukagakusho, and the Max Planck Society. Funding of the project was provided by Anna och Allan L\"ofberg's stipendiefond.
 
%\end{acknowledgements}

{}

\end{document}